\documentclass[twocolumn,superscriptaddress,floatfix]{revtex4-1}
\usepackage{amsfonts,amsmath,graphicx,amssymb,bm}
\usepackage[pdftex,plainpages=false,colorlinks=true,linkcolor=blue, citecolor=blue, urlcolor=blue]{hyperref}
\pdfoutput=2

\usepackage[charter]{mathdesign}

\usepackage[x11names,svgnames]{xcolor}

\newcommand\ev{\mathbf{e}}
\newcommand\av{\mathbf{a}}
\newcommand\rv{\mathbf{r}}
\newcommand\Kv{\mathbf{K}}
\newcommand\kv{\mathbf{k}}
\newcommand\Qv{\mathbf{Q}}

\newcommand\Sigmav{\bm{\Sigma}}

\newcommand\Gv{\mathbf{G}}
\newcommand\Tr{\mathrm{Tr}}
\newcommand\s{\sigma}
\newcommand\pipK{$p$+$ip$-K}
\newcommand\AFpipK{AF$\oplus p$+$ip$-K}
\newcommand\ffrac[2]{{\textstyle\frac{#1}{#2}}}

\begin{document}

%=============================================================================================
\title{Kekul\'e superconductivity and antiferromagnetism on the graphene lattice}

%=============================================================================================
\author{J. P. L. Faye}
\affiliation{D\'epartement de physique and RQMP, Universit\'e de Sherbrooke, Sherbrooke, Qu\'ebec, Canada J1K 2R1}

\author{M. N. Diarra}
\affiliation{D\'epartement de physique, Facult\'e des sciences et techniques, Universit\'e Cheikh Anta Diop de Dakar, Dakar, S\'en\'egal, B.P. 5005 Dakar-Fann }
\author{D. S\'en\'echal}
\affiliation{D\'epartement de physique and RQMP, Universit\'e de Sherbrooke, Sherbrooke, Qu\'ebec, Canada J1K 2R1}
\date{\today}

%=============================================================================================
\begin{abstract}
We investigate superconducting order in the extended Hubbard model on the two-dimensional graphene lattice using the variational cluster approximation (VCA) with an exact diagonalization solver at zero temperature.
Building on the results of Ref.~\cite{Faye:2014}, which identified triplet $p$- and $p+ip$-wave superconductivity as the most favored pairing symmetries in that model, we place uniform SC solutions in competition with a nonuniform Kekul\'e (\pipK) superconducting pattern, similar to those proposed in Ref.~\cite{Roy:2010kq}. We find that the \pipK\ solution is in fact the most favored pairing in most of the phase diagram. In addition, we show that antiferromagnetism can co-exist with the \pipK\ state and that both orders are enhanced by their coexistence.
\end{abstract}
\maketitle

%===============================================================================
\section{Introduction} \label{Sec1}

There is now a considerable body of theoretical work on the possibility that electron-electron interactions in graphene or related materials could lead to superconductivity, at least away from half-filling.
Whereas most studies have focused on singlet, and in particular chiral, $d$-wave superconducivity~\cite{Uchoa:2007,Pathak:2010,Ma:2011,Kiesel:2012fk,Nandkishore:2012fj,Nandkishore:2012kx,Wu:2013}, it has been argued recently that triplet superconductivity would be favored~\cite{Roy:2010kq,Faye:2014}.
Interestingly, certain vortex excitations in triplet, $p+ip$ superconductors have zero-energy Majorana modes in their cores~\cite{Kopnin:1991} and this endows these vortices with non-Abelian statistics~\cite{Ivanov:2001}.

The present authors have proposed in Ref.~\cite{Faye:2014} that triplet superconductivity occurs in the extended Hubbard model defined on the graphene lattice.
These conclusions were based on computations within the Variational Cluster Approximation (VCA) and cellular dynamical mean-field theory (CDMFT).
Many types of superconducting order, both singlet and triplet, were compared in the VCA (that approach provides an approximation to the condensation energy) the conclusion being that chiral, $p+ip$ superconductivity is favored over the non-chiral solution at large enough values of $U$ or $V$.

However, the comparisons performed in \cite{Faye:2014} did not include the possibility of non-uniform superconducting states, in particular along the lines of those proposed by Roy and Herbut~\cite{Roy:2010kq}.
In this work, we investigate that possibility, again using the Variational Cluster Approximation.
We find that the nonuniform, triplet and chiral Kekul\'e superconductivity (\pipK) is favored over the uniform solution ($p+ip$) found in \cite{Faye:2014}, except close to half-filling and for large values of the on-site interaction $U$.
The proposed \pipK\ pairing occurs between electrons of the same valley, and thus Cooper pairs have a non-zero center-of-mass momentum, whereas the pairings studied in Ref.~\cite{Faye:2014} and most other works occur between electrons of opposite valleys.
Note that Kekul\'e pairing was also studied in Ref.~\cite{Black-Schaffer:2014uq}, but for singlet pairing only, and in Ref.~\cite{Kunst2015} in the mean-field approximation and the low-energy limit. In the latter work it was also found to dominate the non-Kekul\'e pairing.

We also investigate the competition between N\'eel antiferromagnetism and \pipK\ superconductivity.
We find that the two phases coexist in a homogeneous manner in some doping range, provided $U$ is large enough.
Coexistence enhances both phases, as if they were in co-operation rather than in competition.

This paper is organized as follows: In Sect.~\ref{sec:model}, we present the model and notation, we quickly review the VCA and we define the various superconducting orders studied.
In Sect.~\ref{sec:results}, we present the results of VCA computations, both for pure superconductivity and in coexistence with antiferromagnetism. We conclude in Sect.~\ref{sec:conclusion}.

%===============================================================================
\section{Model and method} \label{sec:model}

%-------------------------------------------------------------------------------
\subsection{The model Hamiltonian}

The Hamiltonian of the extended Hubbard model on the honeycomb lattice is:
\begin{multline}
\label{eq:model}
H=- t\sum_{\rv\in A,\sigma,j}\left( c^\dagger_{\rv,\sigma}c_{\rv+\ev_j,\sigma} + \mathrm{H.c}\right) -\mu\sum_\rv n_{\rv} \\ 
+ U \sum_{\rv} n_{\rv,\uparrow} n_{\rv,\downarrow} + V \sum_{\rv\in A,j} n_{\rv}n_{\rv+\ev_j}
\end{multline}
where $t$ is the amplitude of the nearest-neighbor hopping matrix (taken as a unit of measure for energy, i.e. $t=1$, in the remainder of this work); $c_{\rv,\sigma}$ ($c^{\dagger}_{\rv,\sigma}$) destroys (resp. creates) an electron of spin $\s$ in a Wannier orbital centered at site $\rv$; $n_{\rv,\s} = c^{\dagger}_{\rv,\sigma}c_{\rv,\sigma}$ is the number of electrons of spin $\s$ at site $\rv$, and $n_\rv=n_{\rv,\uparrow}+n_{\rv,\downarrow}$.
The three vectors $\ev_{1,2,3}$ link a site of sublattice A with its three nearest neighbors on sublattice B, and are oriented at $120^\circ$ of each other.
$\mu$ is the chemical potential, $U$ the on-site repulsion and $V$ the Coulomb interaction between electrons located on nearest-neighbor sites.

On the $U$-$V$ plane and at half-filling, model \eqref{eq:model} has antiferromagnetic (AF) and charge-density-wave phases separated by a normal phase~\cite{Faye:2014}. 
The parameters appropriate for graphene sheets~\cite{Wehling:2011} should fall in the latter phase, in the approximation where interactions beyond nearest-neighbor sites are neglected.
At large $U$, the model can be used for other materials, for instance the compound $\mathrm{In}_3\mathrm{Cu}_2\mathrm{V}\mathrm{O}_9$, which is known from experiments to be antiferromagnetic~\cite{Yan:2012}.

%...............................................................................
\begin{figure}
\centerline{\includegraphics[scale=0.75]{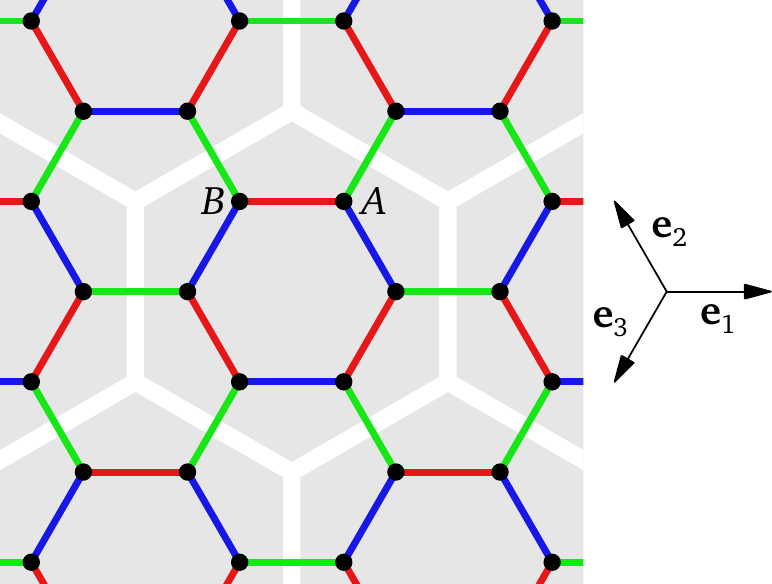}}
\caption{(Color online) Tiling of the honeycomb lattice by 6-site clusters (gray shading) used in VCA.
The three colors (red, green, blue) of the links correspond to the three different phases of the bond triplet pairing in the Kekul\'e pattern defined as \pipK\ in the text.
The A and B sublattices are indicated, as well as the three elementary vectors $\ev_{1,2,3}$.}
\label{fig:hexa}
\end{figure}
%...............................................................................

%-------------------------------------------------------------------------------
\subsection{The variational cluster approximation}

In order to probe the possibility of superconductivity in model \eqref{eq:model}, we use the variational cluster approximation (VCA) with an exact diagonalization solver at zero temperature~\cite{Dahnken:2004}.
This method has been applied to many strongly correlated systems in connection with various broken symmetry phases, in particular superconductivity~\cite{Senechal:2005,Aichhorn:2006}.
For a detailed review of the method, see Refs~\cite{Potthoff:2012fk,Potthoff:2014rt}.

Like other quantum cluster methods, VCA starts by a tiling of the lattice into an infinite number of (usually identical) clusters. 
In this work, we will use the 6-site cluster shown in Fig.~\ref{fig:hexa}.
In VCA, one considers two systems: the original system described by the Hamiltonian $H$, defined on the infinite lattice, and the {\it reference system}, governed by the Hamiltonian $H'$, defined on the cluster only, with the same interaction part as $H$.
Typically, $H'$ will be a restriction of $H$ to the cluster (i.e., with inter-cluster hopping removed), to which various Weiss fields may be added in order to probe broken symmetries.
More generally, any one-body term can be added to $H'$.
The size of the cluster should be small enough for the electron Green function to be computed numerically.

The optimal one-body part of $H'$ is determined by a variational principle. More precisely, the electron self-energy $\Sigmav$ associated with $H'$ is used as a variational self-energy, in order to construct the following Potthoff self-energy functional~\cite{Potthoff:2003b}:
\begin{multline}\label{eq:omega}
\Omega[\Sigmav(\xi)]=\Omega'[\Sigmav(\xi)]\\ +\Tr\ln[-(\Gv^{-1}_0 -\Sigmav(\xi))^{-1}]-\Tr\ln(-\Gv'(\xi))
\end{multline} 
The quantities $\Gv'$ and $\Gv_0$ above are respectively the physical Green function of the cluster and the non-interacting Green function of the lattice. 
The symbol $\xi$ stands for a small collection of parameters that define the one-body part of $H'$. 
$\Tr$ is a functional trace, i.e. a sum over frequencies, momenta and bands, and $\Omega'$ is the grand potential of the cluster, i.e. its ground state energy, since the chemical potential $\mu$ is included in the Hamiltonian.
$\Gv'(\omega)$ and $\Omega'$ are computed numerically via the Lanczos method at zero temperature.

The Potthoff functional $\Omega[\Sigmav(\xi)]$ in Eq.~\eqref{eq:omega} is computed exactly, but on a restricted space of the self-energies $\Sigmav(\xi)$ that are the physical self-energies of the reference Hamiltonian $H'$.
We use a standard optimization method (e.g. Newton-Raphson) in the space of parameters $\xi$ to find the stationary value of $\Omega(\xi)$:
\begin{equation}\label{eq:Euler}
\frac{\partial\Omega(\xi)}{\partial\xi} = 0
\end{equation}
This represents the best possible value of the self-energy $\Sigmav$, which is used, together with the non-interacting Green function $\Gv_0$, to construct an approximate Green function $\Gv$ for the original lattice Hamiltonian $H$.
From that Green function one can compute the average of any one-body operator, in particular order parameters associated with antiferromagnetism or superconductivity.
The actual value of $\Omega(\xi)$ at the stationary point is a good approximation to the physical grand potential of the lattice Hamiltonian $H$.

There may be more than one stationary solutions to Eq.~\eqref{eq:Euler}. For instance: A {\it normal state} solution in which all Weiss fields used to describe broken symmetries are zero, and another solution, with a non-zero Weiss field, describing a broken symmetry state. 
As an additional principle, we assert that the solution with the lowest value of the functional~\eqref{eq:omega} is the physical solution~\cite{Potthoff:2006mb}.
Thus competing phases may be compared via their value of the grand potential $\Omega$, obtained by introducing different Weiss fields. 

A formal requirement of the method is that the interaction terms are the same in $H$ and $H'$.
This cannot be true if extended interactions are included, because the tiling of the lattice into clusters breaks inter-cluster interactions.
In that case, an additional approximation is needed: the dynamical Hartree approximation~\cite{Senechal:2013bh,Faye:2014}, in which the extended interaction term in \eqref{eq:model} is replaced by \begin{multline}\label{eq:3}
V \sum_{\rv,\rv'} n_{\rv}n_{\rv'} = \frac12\sum_{\rv,\rv'} V_{\rv,\rv'}^\mathrm{c} n_\rv n_{\rv'} \\ + 
\frac12\sum_{\rv,\rv'} V_{\rv,\rv'}^\mathrm{ic} (\bar n_\rv n_{\rv'} + n_\rv\bar n_{\rv'} - \bar n_\rv \bar n_{\rv'})
\end{multline}
where $V_{\rv,\rv'}^\mathrm{c}$ denotes the nearest-neighbor interactions between sites belonging to the same cluster and $V_{\rv,\rv'}^\mathrm{ic}$ the interactions between sites belonging to different clusters.
In this notation the positions $\rv,\rv'$ may be restricted the cluster, as they are folded back appropriately onto a single cluster.
The mean fields $\bar n_\rv$ must be determined self-consistently via a repeated application of the VCA method.
For the 6-site cluster used in this work, if we ignore the possibility of charge order, this mean-field decomposition amounts to shifting the chemical potential. 
Thus the one-body operator added to each cluster is
\begin{equation}\label{eq:mean_fields}
V^\mathrm{mf} = V n\hat N - \ffrac12\cdot 6 V n^2
\end{equation}
where $n$ is the average occupation of each site and $\hat N$ is the total electron number operator on the cluster.

%-------------------------------------------------------------------------------
\subsection{Superconductivity}

The VCA is a real-space method with an emphasis on short-range correlations, because of the small size of the clusters.
In order to probe broken symmetries, one therefore introduces a set of {\it local} Weiss fields.
In particular, for superconductivity, we introduce the following nearest-neighbor, singlet and triplet pairing operators:
\begin{equation}
\label{eq:pairing}
\begin{aligned}
\hat{\Delta}^{S}_{\rv,i} &= c_{\rv,\uparrow}c_{\rv+\ev_i,\downarrow} - c_{\rv,\downarrow}c_{\rv+\ev_i,\uparrow} \\
\hat{\Delta}^{T}_{\rv,i} &= c_{\rv,\uparrow}c_{\rv+\ev_i,\downarrow} + c_{\rv,\downarrow}c_{\rv+\ev_i,\uparrow}
\end{aligned}
\end{equation}
where the nearest-neighbor vectors $\ev_i$ are defined on Fig.~\ref{fig:hexa}.
We will only consider the $z$-component of the triplet operator. The other two
components ($x$ and $y$) would be 
\begin{equation}
\begin{aligned}
\hat{\Delta}^{T}_{\rv,i,x} &= 
-c_{\rv,\uparrow}c_{\rv+\ev_i,\uparrow} + c_{\rv,\downarrow}c_{\rv+\ev_i,\downarrow}\\
\hat{\Delta}^{T}_{\rv,i,y} &= 
-i c_{\rv,\uparrow}c_{\rv+\ev_i,\uparrow} - i c_{\rv,\downarrow}c_{\rv+\ev_i,\downarrow}\\
\end{aligned}
\end{equation}
We will assume rotational invariance, so in principle this restriction to the $z$ component of the triplet field is not a problem.

%~~~~~~~~~~~~~~~~~~~~~~~~~~~~~~~~~~~~~~~~~~~~~~~~~~~~~~~~~~~~~~~~~~~~~~~~~~~~~~~
\begin{table}
\caption{Irreducible representations (irreps) of $D_{6h}$ associated with the six pairing operators defined on nearest-neighbor sites. The phases refer to Eq.~\eqref{eq:OP} and the symmetries may be understood by considering
Fig.~2 of Ref.~\cite{Faye:2014}.
\label{table:irreps}}
\begin{tabular}{llll}
\hline\hline
Irrep~~ & symbol~~ & & $(\phi_1, \phi_2, \phi_3)$ \\[3pt]
\hline
$A_1$ & $s$ & singlet~~ & $(1, 1, 1)$ \\
$B_1$ & $f$ & triplet & $(1, 1, 1)$ \\
$E_1$ & $d$ & singlet & $(1, -1, 0)$ and $(0, 1, -1)$ \\
$E_2$ & $p$ & triplet & $(1, -1, 0)$ and $(0, 1, -1)$ \\
\multicolumn{4}{c}{chiral representations} \\
$E_1$& $d+id$ & singlet & $(1, e^{2\pi i/3}, e^{4\pi i/3})$ \\
& $d-id$ & singlet & $(1, e^{-2\pi i/3}, e^{-4\pi i/3})$ \\
$E_2$& $p+ip$ & triplet & $(1, e^{2\pi i/3}, e^{4\pi i/3})$ \\
& $p-ip$ & triplet & $(1, e^{-2\pi i/3}, e^{-4\pi i/3})$ \\
\hline\hline
\end{tabular}
\end{table}
%~~~~~~~~~~~~~~~~~~~~~~~~~~~~~~~~~~~~~~~~~~~~~~~~~~~~~~~~~~~~~~~~~~~~~~~~~~~~~~~

From the elementary operators \eqref{eq:pairing}, one can define lattice-wide pairing operators as follows:
\begin{equation}\label{eq:OP}
\hat{\Delta}^{S(T)}_\Qv = \sum_ {\rv\in A,j}\left(\hat{\Delta}^{S(T)}_{\rv,j}e^{i(\Qv\cdot\rv +\phi_j)} + \mathrm{H.c.} \right)
\end{equation}
where the three phases $\phi_j$ and the wave vector $\Qv$ define the precise superconducting order under study.
In Ref.~\cite{Faye:2014} we considered all the possibilities listed in Table~\ref{table:irreps}, at $\Qv=0$.
The conclusion was that the chiral $p+ip$ order was favored in most of the phase diagram explored.

In this work, we study the same list of representations, this time at $\Qv=\Kv=(2\pi/3,2\pi/3\sqrt3)$ (the Dirac wave vector).
Pairing patterns associated with this wave vector will be qualified as Kekul\'e patterns, following Ref.~\cite{Roy:2010kq}, and will be noted as $s$-K, $d$-K, \pipK, etc.
 
What is the meaning of this wave vector for pairing? This is made clear by going to a $k$-space, rather than a real-space, description of the pairing operator.
It is then important to distinguish between annihilation operators for sublattices A and B:
\begin{equation}
c^{A,B}_{\rv,\s} = \frac1{\sqrt{N}}\sum_\kv e^{-i\kv\cdot\rv} c^{A,B}_{\kv,\s}
\end{equation}
where now $\rv$ only refers to a Bravais lattice position, say on sublattice $A$.
Let $\av_j$ ($j=1,2,3$) stand for the Bravais lattice vectors associated with the $B$ sublattice nearest neighbors of a given site on sublattice $A$; in other words: $\av_j = \ev_j - \ev_1$. Then one shows simply that the pairing Hamiltonian \eqref{eq:OP} is
\begin{equation}\label{eq:OPk}
\hat{\Delta}^{(T,S)}_\Qv = \sum_{\kv,j} e^{i(\kv-\Qv)\cdot\av_j + i\phi_j}
\left( c^A_{\kv\uparrow}c^B_{\Qv-\kv\downarrow} \pm c^A_{\kv\downarrow}c^B_{\Qv-\kv\uparrow} \right)
+ \mathrm{H.c.}~
\end{equation}
where the relative sign of the two terms is $-1$ in the case of singlet pairing.

%...............................................................................
\begin{figure*}
\begin{minipage}[c]{0.64\textwidth}
\includegraphics[width=\hsize]{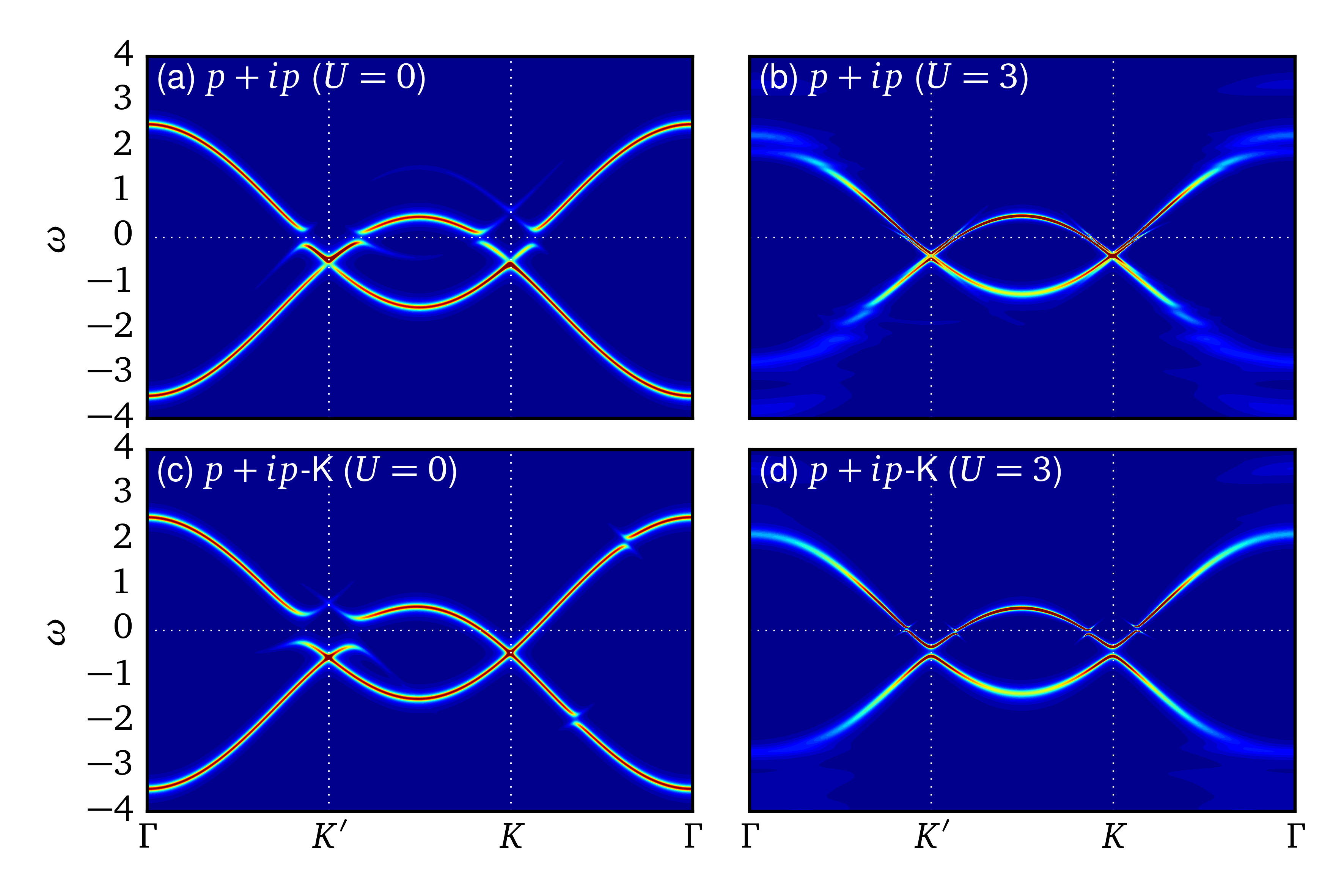}
\end{minipage}
\begin{minipage}[c]{0.35\textwidth}
\caption{(Color online) Spectral function of BCS-like ($U=0$) solutions (left) and interacting, VCA solutions (right) for $p+ip$ (top) and \pipK\ (bottom) pairing. Doping is set at 5\%. The BCS solutions were giving pairing amplitudes $\Delta=0.1$, but this exaggerates the order parameter compared to the VCA solutions. Notice that the \pipK\ state is gapless (hidden) at $U=0$ but has a gap in the presence of interactions, whereas the opposite is true of the $p+ip$ state. Also, the asymmetry between the two distinct Dirac points $\Kv$ and $\Kv'$ is very strong in the non-interacting case, but much weaker in the presence of interactions, to the extent that it is barely visible here. A fixed Lorenzian broadening $\eta=0.05$ has been used at $U=0$, whereas $\eta$ increases linearly from $0.03$ to $0.2$ as a function of $|\omega|$ in the interacting case.
}
\label{fig:dispersion}
\end{minipage}
\end{figure*}
%...............................................................................

We see from this expression that pairing occurs between electrons of opposite valleys ($\kv$ and $-\kv$) when $\Qv=0$. On the other hand, if $\Qv=\Kv$, the momentum of the pair is $\Kv$, which is equivalent to $-2\Kv \sim 2\Kv'$ modulo a reciprocal lattice vector. This corresponds to pairing between electrons within a single valley (the $\Kv'$ valley if $\Qv=\Kv$ and vice-versa).
Time reversal of \eqref{eq:OPk} amounts to changing the sign of the phases $\phi_j$ {\em and} going from $\Qv=\Kv$ to $\Qv = -\Kv \sim \Kv'$. 

The particular order \pipK\ will turn out to be dominant. In that pattern, the phases of the pairing operators on each link follow the pattern illustrated on Fig.~\ref{fig:hexa}: The three colors correspond to phases separated by $2\pi/3$.
The time-reversed state has the same phase pattern, except that the order of colors is reversed as one goes clockwise around a site.

For each type of superconducting order found, we can compute the condensation energy
\begin{equation}
E^\mathrm{cond.} = E^\mathrm{norm.} - E^\mathrm{sc}~,
\end{equation}
i.e., the difference between the normal state energy, found by varying only $\mu'$, and the superconducting energy, found by varying both $\mu'$ and $\Delta$. The ground state energy of any solution is given by $E = \Omega + \mu n$, where $\Omega$, the grand potential, is approximated by the value of the Potthoff functional \eqref{eq:omega} at the appropriate values of $(\mu',\Delta)$.
When many competing broken symmetry solutions are obtained, the one with the largest condensation energy is favored.

The left panels of Fig.~\ref{fig:dispersion} show the spectral function of the quasiparticles in the non-interacting Hamiltonian $H_0+\Delta 
\hat{\Delta}^{T}_\Qv$ for $\Delta=0.1$, 5\% doping, and $\Qv=0$ (top) and $\Qv=\Kv$ (bottom), i.e., respectively in the $p+ip$ and the \pipK\ solutions.
As we see, the former has a superconducting gap, but the latter doesn't. Thus, in this sense, \pipK\ superconductivity is hidden~\cite{Uchoa:2007}.
However, as shown on the right panels of the same figure and as explained below, the opposite is true in the presence of interactions, when the solutions are obtained from VCA.
In both types of pairing, there is an asymmetry between the two Dirac points $\Kv$ and $\Kv'$, which is natural given that both solutions break time-reversal invariance, and that $\Kv$ is mapped onto $\Kv'$ by time reversal. This asymmetry is much stronger for \pipK\ pairing, since in that case the electrons involved in the pairing come from one of the valleys only ($\Kv'$ in the case illustrated on the figure).

%...............................................................................
\begin{figure}[b]
\centerline{\includegraphics[width=\hsize]{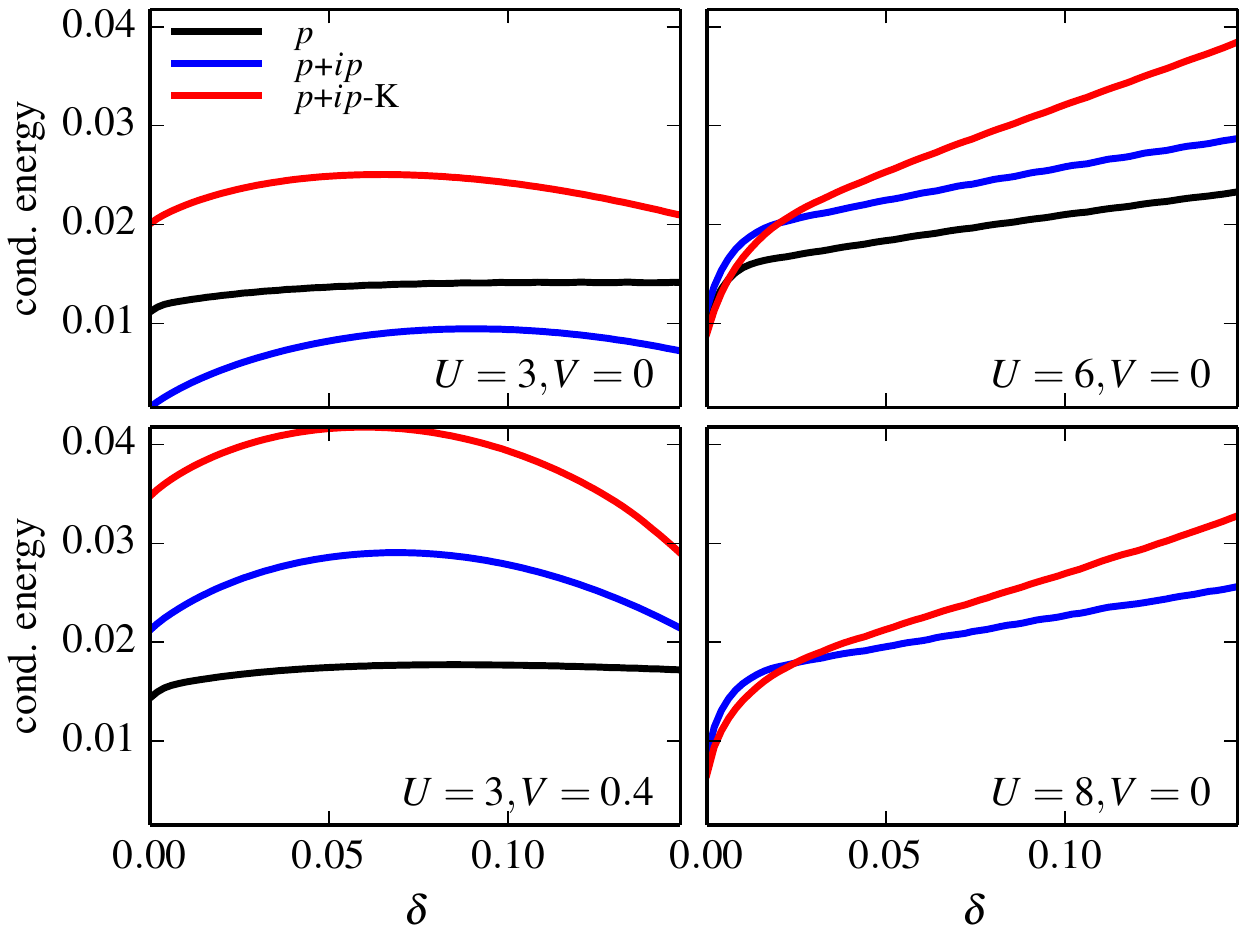}}
\caption{(Color online) Condensation energy as a function of doping $\delta=1-n$ at various values of $U$ and $V$, for $p$, $p+ip$ and \pipK\ superconductivity.}
\label{CE-U3-V}
\end{figure}
%...............................................................................

%===============================================================================
\section{Results and discussion}\label{sec:results}

%-------------------------------------------------------------------------------
\subsection{Pure superconductivity}

Now, we present our results obtained by the VCA method, starting with pure superconductivity, i.e., ignoring any other symmetry breaking.
The competition between superconductivity and antiferromagnetism will be addressed in the next subsection.

In Fig.~\ref{CE-U3-V}, we show the condensation energies at $U=3t$ and several values of the extended interaction $V$.
The singlet paring ($s$, $d$ and $d\pm id$) and the $f$-wave (triplet) pairing were also investigated, both for $\Qv=0$ and $\Qv=\Kv$, but their condensation energies are either negative or very small compared with the $p$-wave solutions and therefore are not shown.
The $p$ and $p+ip$ solutions are borrowed from Ref.~\cite{Faye:2014}.
At $U=3$, the condensation energy of the \pipK\ solution is higher in the range studied, i.e., this solution is favored compared to all other possibilities, for all values of $V$ studied.
Note that $U=3t$ is very close to the appropriate value for graphene $(U\approx 3.3t)$ \cite{Wehling:2011}.
However, the corresponding value of $V$ ($\approx 2t$) would lead, in our approximation, to a charge-density wave at half-filling, because of our neglect of longer-range Coulomb interactions.
Therefore we have not pushed $V$ that far.
At larger values of $U$, the \pipK\ phase is still dominant, except at small doping, where the $p+ip$ solution is favored.

The right panels of Fig.~\ref{fig:dispersion} show the spectral function of the $p+ip$ (top) and \pipK\ (bottom) solutions obtained from VCA at $U=3$ and 5\% doping.
A striking feature is the closure of the superconducting gap of the $p+ip$ solution, and the opening of the gap at $\Kv$ in the \pipK\ solution, absent from the non-interacting (BCS) solution.
It may be tempting to conclude that this phenomenon explains why the \pipK\ is energetically favored in VCA, but one must be careful since the energetics of VCA is subtle and does not simply involve the filling of quasiparticle levels. The asymmetry between the two Dirac points, strong in the non-interacting case, is much attenuated by interactions. Yet it has not completely disappeared, although this is not visible in Fig.~\ref{fig:dispersion}. In particular, it can be shown that the total spectral weight at the Fermi level is still larger around one of the Dirac points than around the other.

Fig.~\ref{fig:U-delta-AF} shows an approximate phase diagram at $V=0$ on the $U$-$\delta$ plane, where the transition from \pipK\ to $p+ip$ is indicated in blue.
However, as we will see below, when N\'eel antiferromagnetism is allowed, the pure $p+ip$ solution has a higher energy than a solution where antiferromagnetism and \pipK\ order coexist.

\begin{figure}
\centerline{\includegraphics[width=7.5cm]{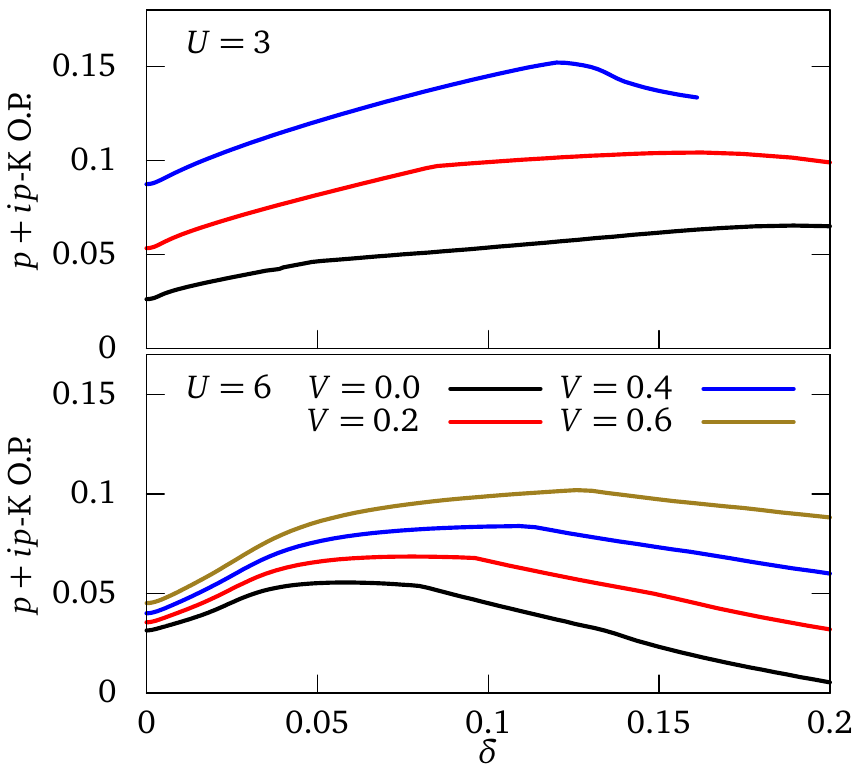}}
\caption{(Color online) \pipK\ order parameter as a function of doping at $U=3t$ (top panel) and $U = 6t$ (bottom panel), and several values of $V$.}
\label{OP-U3-U6}
\end{figure}

In VCA, we define the order parameter as the expectation value $\langle\hat\Delta\rangle$ of the operator $\hat\Delta$ defined in Eq.~\eqref{eq:OP}.
It is computed from the approximate Green function $\Gv$ found from the VCA self-energy.
Fig.~\ref{OP-U3-U6} shows how the chiral superconducting order parameter $p+ip$-K varies as a function of doping $\delta$ at $U=3t$ and $U = 6t$, for several values of $V$. 
Notice the non-zero value of the order parameter at half-filling.
The amplitude of the order parameter increases with $V$ and has a broad dome-like shape.

%-------------------------------------------------------------------------------
\subsection{Superconductivity and Antiferromagnetism}\label{sec:SC-AF}

It is commonly accepted that the half-filled Hubbard model on the graphene lattice with nearest-neighbor hopping has an AF ground state beyond a certain value of the on-site interaction $U$ at $V=0$~\cite{Furukawa:2001qf,Honerkamp:2008qd,Raghu:2010fk,Sorella:2012qf,Hassan:2013uq,Assaad:2013fk,Arya:2015fk}.
That critical value is estimated to be $U_c\sim 3.86$~\cite{Sorella:2012qf}.
In Ref.~\cite{Faye:2014}, the same critical value was obtained in VCA, and in addition its dependence on $V$ was mapped out.

We now place this AF solution in competition with the \pipK\ superconducting solution, by letting the corresponding Weiss fields vary simultaneously, first at half-filling (as a function of $U$), and then away from half-filling for two values of $U$. We will set $V=0$ in this section.

%...............................................................................
%...............................................................................

%...............................................................................
\begin{figure}[h]
\centerline{\includegraphics[width=7.5cm]{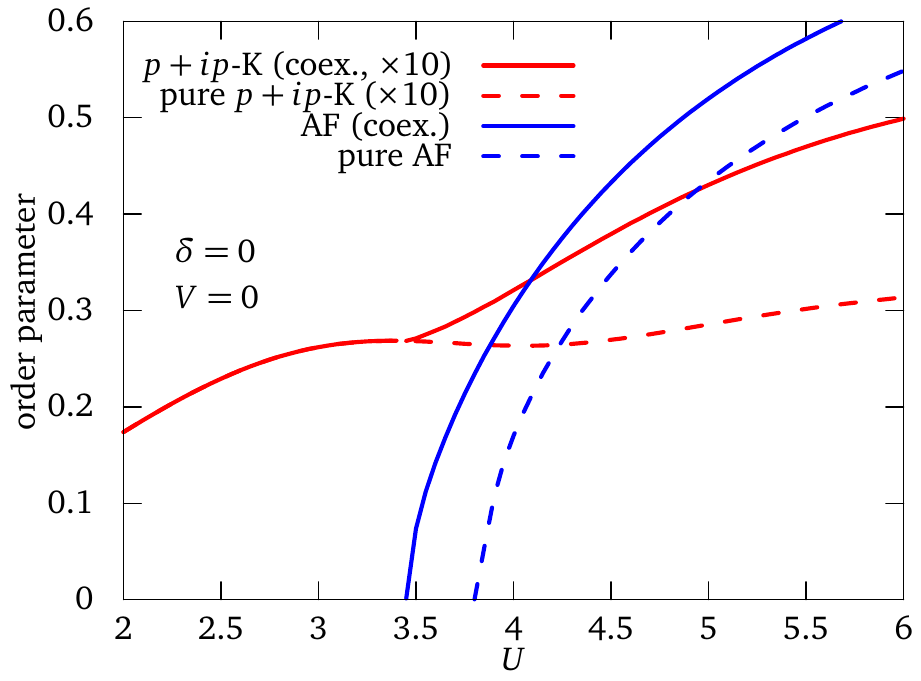}}
\caption{(Color online) Superconducting \pipK\ and AF order parameters as a function of the on-site repulsion $U$ at half-filling.
The dashed curves are the pure solutions and the full curves the coexistence solutions.}
\label{AF-HF}
\end{figure}
%...............................................................................

In VCA, N\'eel antiferromagnetism is probed by adding the following term to the Hamiltonian $H'$ of the reference system:
\begin{equation}
H'_{\rm AF} = h\left\{ \sum_{i\in A} (n_{i\uparrow}-n_{i\downarrow}) - \sum_{i\in B} (n_{i\uparrow}-n_{i\downarrow}) \right\}
\end{equation}
where $h$ is the Weiss field to be treated as a variational parameter in the same way as $\Delta$ and $\mu'$.
At half-filling, the value of $\mu'$ is fixed to $\mu=U/2$ by particle-hole symmetry, thus we need only vary $\Delta$ and $h$.
We find solutions to Eq.~\eqref{eq:Euler} where $\Delta=0$ and $h\ne 0$ (pure antiferromagnetism), $\Delta\ne0$ and $h=0$ (pure \pipK\ superconductivity, like above), and solutions with $\Delta\ne0$ and $h\ne0$, corresponding to a microscopic coexistence of antiferromagnetism and superconductivity (\AFpipK).
Note that the presence of antiferromagnetism breaks rotation invariance in spin space, and thus the three components of the triplet pairing are in principle no longer degenerate. We have carried computations only for the $z$-component defined in Eq.~\eqref{eq:pairing}, assuming that this is the most favored component, as it is also the direction chosen for antiferromagnetism.

On Fig.~\ref{AF-HF}, we show the individual AF and the triplet \pipK\ order parameters in the pure and in the coexistence solutions.
Interestingly, there seems to be co-operation between the two orders instead of competition, as their amplitudes in the coexistence solution is larger than in the pure solutions.

\begin{figure}[h]
\centerline{\includegraphics[width=7.5cm]{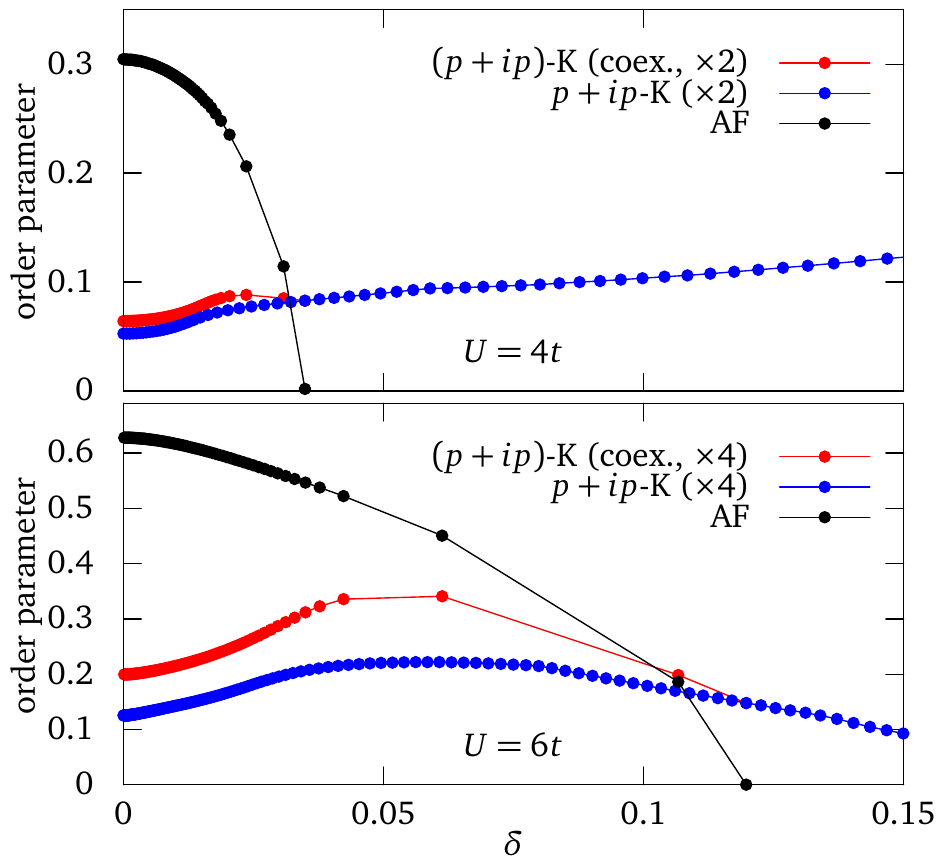}}
\caption{(Color online) Superconducting order parameters \pipK\ and antiferromagnetic (AF) versus doping at $U=4t$ and $U=6t$ and $V=0$ Top and down panel respectively.
At $U=4t$ the antiferromagnetism disappears around $3\%$ of doping and $10\%$ at $U=6t$.
The doted blue curve is the pure superconducting order parameter, i.e. when the Weiss field for antiferromagnetism $h_\mathrm{AF} = 0$ and the red curve corresponds to the coexistence solution with AF.
Notice the increase of the amplitude of the superconducting order parameter in presence of AF.}
\label{AF-U4-U6}
\end{figure}

Fig.~\ref{AF-U4-U6} (top) shows the \pipK\ and AF order parameters as a function of doping $\delta$ at $U=4t$ and $V=0$, very close to the critical value for the onset of antiferromagnetism in the absence of superconductivity ($U_c = 3.82t$; see also Fig.~\ref{AF-HF}).
The blue curve is the pure \pipK\ order parameter, the red curve is the \pipK\ order parameter in the coexistence solution and the black curve is the AF order parameter, also in the coexistence solution.
The bottom panel shows similar data, but at $U =6t$.
The extent of antiferromagnetism with doping increases with $U$: 3\% at $U=4$ vs 12\% at $U=6$.
Again, \pipK\ (triplet) superconductivity is enhanced by its coexistence with antiferromagnetism.
On the other hand, $d$-wave (singlet) superconductivity on the square-lattice Hubbard model is partially suppressed by its coexistence with antiferromagnetism~\cite{Senechal:2005,Aichhorn:2006,Kancharla:2008vn}. 

%...............................................................................
%...............................................................................
%...............................................................................
\begin{figure}
\centerline{\includegraphics[scale=0.85]{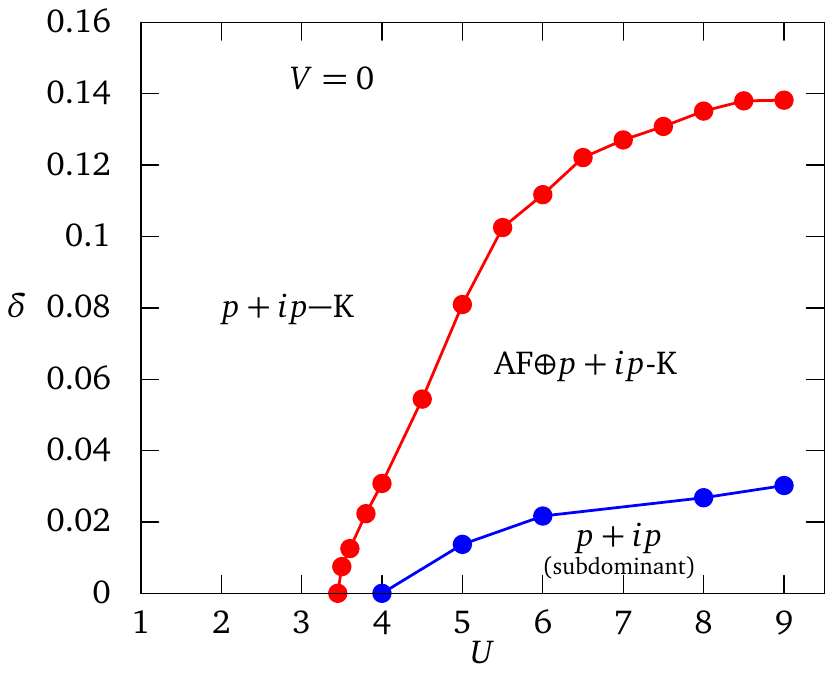}}
\caption{(Color online) Phase diagram on the $U$-$\delta$ plane at $V=0$.
The red curve corresponds to the critical line where the antiferromagnetic (AF) solution that co-exists  with  the  chiral \pipK\ solution disappears. 
In the region below the blue curve, the uniform $p+ip$ solution is preferred over the \pipK\ solution when antiferromagnetism is suppressed, but the \AFpipK\ solution has lower energy, and therefore dominates in that region.}
\label{fig:U-delta-AF}
\end{figure}
%...............................................................................

In Fig.~\ref{fig:U-delta-AF}, we show the phase diagram obtained by collecting the critical doping values where the AF order parameter goes to zero for each value of $U$.
At strong coupling and moderate doping, the \AFpipK\ solution has the lowest energy, whereas the pure \pipK\ state dominates in the rest of the phase diagram. 
The region that was occupied by the pure $p+ip$ solution when neglecting N\'eel antiferromagnetism (under the blue curve) is now occupied by the \AFpipK\ solution. 
We also investigated  the possibility of homogeneous coexistence between the uniform $p+ip$ and AF solutions.
For $U < 9.2$, there is no coexistence between these solutions. 
On the other hand, for $U > 9.2$ and only at half-filling, $p+ip$ superconductivity co-exists with N\'eel AF and becomes the best solution. 
However, this coexistence solution is destroyed at very small doping ($\delta \sim 0.2 \% $ for $U = 10$). 

We have checked that at large $U$, the effect of $V$ is not very important and therefore we do not expect the phase diagram to change much as a function of $V$.
If we compare with the results of Ref.~\cite{Faye:2014}, it becomes clear that the regions which were believed to be occupied by the uniform solutions of type $p$ or $p+ip$ are instead either occupied by the \pipK\ or the \AFpipK\ solution.

%===============================================================================
\section{Conclusion} \label{sec:conclusion}

Using the Variational Cluster Approximation (VCA), we investigated possible superconducting phases in the extended Hubbard model on the graphene lattice.
We went beyond the uniform states studied in Ref.~\cite{Faye:2014} by comparing with the so-called Kekul\'e superconducting pattern (\pipK) introduced in Ref.~\cite{Roy:2010kq}, which breaks translation invariance.
We found that the \pipK\ phase or the \AFpipK\ coexistence phase have a lower energy in most of the phase diagram, except at high values of the on-site repulsion $U$ and at very low doping, where a coexistence between antiferromagnetism and the uniform chiral triplet $p+ip$ solution is preferred.
Homogeneous coexistence of N\'eel antiferromagnetism and \pipK\ superconductivity occurs beyond a critical value of $U$ and below a critical doping as shown on Fig.~\ref{fig:U-delta-AF}.
We observe that N\'eel antiferromagnetism and \pipK\ superconductivity are enhanced by the possibility of co-existence.
This is completely different from what is observed theoretically on the square lattice, where antiferromagnetism and $d$-wave superconductivity are both negatively affected by their coexistence.

The \pipK\ solution should be experimentally distinguishable from the uniform $p+ip$ solution as it breaks translation invariance: Cooper pairs have a center-of-mass momentum $\Kv$ (or $\Kv'$). Thus, Andreev reflections should behave differently, for instance be more likely when the momentum of the incident electron belongs to a valley rather than its opposite valley. A detailed analysis of this effect still has to be done. Also, as mentioned above, the spectral weight at the Fermi level is larger around one of the Dirac points compared to the other, and this could be seen by angle-resolved photoemission spectroscopy.

Whereas the microscopic parameters adequate for graphene do not lead to N\'eel order, things are different in the planar material $\mathrm{In}_3\mathrm{Cu}_2\mathrm{V}\mathrm{O}_9$, which also crystallizes on the honeycomb lattice~\cite{Yan:2012}.
There, the pure compound is effectively half-filled and has N\'eel order.
Doping is performed by replacing Cu ions by non-magnetic Zn ions and N\'eel order disappears at $\delta\sim 30\%$. 
But in that case the Hubbard model is not entirely appropriate, since doping is obtained by substitution as opposed to charge carrier injection.

%===============================================================================
\begin{acknowledgments}
Computing resources were provided by Compute Canada and Calcul Qu\'ebec.
This research is supported by NSERC grant no RGPIN-2015-05598 (Canada) and by FRQNT (Qu\'ebec).

\end{acknowledgments}
 
%\bibliography{biblio}

%merlin.mbs 2010-03-15 4.21a (PWD, AO, DPC)

\end{document}